\documentclass[elsevier,onecolumn,12 pt,showpacs,showkeys,groupedaddress]{revtex4-1}

\usepackage{amsmath, amssymb}
\usepackage{graphicx}
\usepackage{subfigure}
\usepackage{color}
\newcommand*{\rom}[1]{}
\begin{document}


\title{Long-range s-wave interactions in Bose-Einstein Condensates: An exact correspondence between truncated free energy and dynamics}

\author{Abhijit Pendse}\thanks{abhijeet.pendse@students.iiserpune.ac.in}
\affiliation{Department of Physics, Indian Institute of Science Education and Research, Pune, Maharashtra 411008, India.}

\date{\today}

\begin{abstract}
We consider the Gross-Pitaevskii(GP) model of a Bose-Einstein Condensate(BEC) with non-local s-wave interactions. The non-locality is represented by corrections to the local GP equation. Due to such corrections to the GP equation, there arise corrections to the free energy functional as well. We present here a proof of the exact correspondence between the free energy and the dynamics for typical terms appearing while considering corrections to the GP equation at any order. This non-trivial correspondence can be used to study BECs perturbatively while going beyond the Fermi pseudopotential.
\end{abstract}


\maketitle

\section{Introduction}

The observation of Bose-Einstein Condensation(BEC) in dilute atomic gases was a crucial step in BEC physics\cite{atomicbec_ketterle}. This BEC of interacting bosons has been extensively studied theoretically. The standard approach of studying this interacting BEC is by using the Gross-Pitaevskii(GP) equation\cite{pethick,gp_gross}. In the GP theory, inter-atomic interactions of the BEC are modelled by considering symmetric s-wave scattering between boson pairs. The validity of such an approach has been argued for mainly due to the extremely low temperature and diluteness of the BEC of atomic gases. The theoretical predictions of the GP theory has been verified by experiments in dilute atomic BEC. To mention a few, GP theory has been successful in explaining ground state of a trapped BEC\cite{becreview_leggett}, elementary excitations\cite{excitations_mewes,excitations_stringari}, solitons\cite{solitons_jackson,solitons_burger,solitons_abdullaev}, surface modes\cite{surface_khawaja,surface_onofrio} and vortices\cite{vortex_pitaevskii,vortices_fetter,vortices_lundh} in a BEC.

\par
In the dilute BEC ($a<<n^{-1/3}$)regime, the inter-particle interaction is approximated by a delta-function pseudopotential\cite{ps}. This is due to the fact that in the regime of symmetric interactions, any soft effective pseudopotential $V_{eff}(\bf{r-r^{'}})$ with $\int{{\bf{dr}}\;V_{eff}(\bf{r-r^{'}})}=g$ can be used, where $g=\frac{4\pi\hbar^{2}a}{m}$ gives the strength of interaction ($a$ is the s-wave scattering length). The approximation of the delta-function pseudopotential is a good first order approximation in the low temperature and dilute BEC regime as mentioned earlier. The local GP equation arising from such a pseudopotential considers the energy independence of the s-wave scattering. However, Fu \textit{et al.} have shown that the energy dependence of the s-wave scattering can be taken into account by considering corrections to the local GP equation\cite{nonlocal_wang}. This approach takes into account an effective range expansion due to the non-locality of s-wave scattering. It has been shown, for example in the works of Collin \textit{et al.} and Fu \textit{et al.} that such corrections to the local GP equation would involve derivatives of the form $\psi\nabla^{2}|\psi|^{2}$ and higher even orders\cite{nonlocal_wang, nonlocal_pethick, nonlocal_ketterle}. The immediate question which arises is what would be the corrections to the energy functional due to corrections to the local GP equation? Hence it becomes important to identify the contributions to the energy functional arising from such correction terms. In this paper, we show a rigorous combinatorial proof of the fact that any term of the form $(c|\psi|^{2}/2)\nabla^{l}|\psi|^{2}$ added to the energy functional of the local GP equation would contribute a term $(c\psi)\nabla^{l}|\psi|^{2}$ to the local GP equation, where $l$ is a positive even number. We further show that when $l$ is odd, this correspondence is absent. This makes the correspondence non-trivial and paves the way to study the energetics of corrections on top of the local GP equation. Also, since there exists a term by term correspondence, any number of correction terms can be incorporated on top of the local GP equation.

\par 

The general Gross-Pitaevskii(GP) equation for the complex order parameter $\psi(\bf{r,t})$ of the condensate in the absence of external potential is given as

\begin{equation}
\begin{split}
i\hbar\frac{\partial\psi(\bf{r,t})}{\partial t}&= -\frac{\hbar^{2}}{2m} \nabla^{2} \psi(\bf{r,t})\\
&+ \psi(\bf{r,t}) \int{\bf{dr^{'}} \psi^{*}(\bf{r',t}) V(\bf{r'-r}) \psi(\bf{r',t})} ,
\end{split}
\label{eq:original}
\end{equation}

 which can be derived from the energy functional

\begin{equation}
\begin{split}
E&=\frac{\hbar^{2}}{2m}\int{\bf{dr}\;|\nabla \psi(r)|^{2}}\\
&+\int{\bf{dr}\;\frac{|\psi(r)|^{2}}{2}\int{\bf{dr^{'}}\psi^{*}(\bf{r^{'}}) V(\bf{r-r^{'}})\psi(\bf{r^{'}})}}.
\label{eq:energy}
\end{split}
\end{equation}
The local form of the GP equation is obtained by replacing the interactions between particles with a pseudopotential $\delta(\bf{r-r'})$. This local form of the GP equation is

\begin{equation}
i\hbar\frac{\partial\psi(\bf{r,t})}{\partial t}= -\frac{\hbar ^{2}}{2m} \nabla^{2} \psi(\bf{r,t}) + g\;\psi(\bf{r,t})|\psi(\bf{r,t})|^{2} .
\label{eq:contact}
\end{equation}

The energy functional for the local GP equation can be obtained by a similar replacement of the inter-atomic interactions by the delta function pseudopotential in Eq.(\ref{eq:energy}) giving

\begin{equation*}
E=\frac{\hbar^{2}}{2m}\int{\bf{dr}\;|\nabla \psi(r)|^{2}}+\frac{g}{2}\int{\bf{dr}\;(|\psi(r)|^{2})^{2}}.
\end{equation*}

This local form has been successful in explaining many properties of a BEC. But now if we want to consider the effects due to non-locality of the s-wave scattering interactions, we need to consider corrections to the local GP equation and obtain the corresponding energy functional. Collin \textit{et al.}\cite{nonlocal_pethick} have shown that such corrections to the local GP equation are represented by terms of the form $\alpha\psi\nabla^{l}|\psi|^{2}$ where $l$ is a positive even number. Let us now prove the correspondence between corrections to the local GP equation and the corrections to the energy functional mentioned above, which would enable to study the corrections to the local GP equation.

\vspace{10mm}

\section{\label{sec:proof}Proof of correspondence}
We now propose that any correction(addition) to the local GP equation of the form $\alpha\psi\nabla^{l}|\psi|^{2}$ would correspond to an addition to the energy functional of the form $\alpha\frac{|\psi|^{2}}{2}\nabla^{l}|\psi|^{2}$, where $l$ is even and $\alpha$ is an arbitrary constant. For simplicity, we consider a $\psi$ which is constant along y and z direction, but varies only along the x direction. Hence the laplacian would involve derivatives in $x$ only.

\par
Consider the term in the energy functional $G=\frac{|\psi|^{2}}{2} \partial _{x}^{l}|\psi|^{2}=\frac{|\psi|^{2}}{2}\Big(\sum_{n=0}^{l} {l\choose n} (\partial_{x}^{l-n}\psi)  (\partial_{x}^{n} \psi^{*}) \Big)$. Let us look at the functional derivative of $G$ with respect to $\psi^{*}$.  
\begin{widetext}
\begin{equation*}
\begin{split}
\frac{\delta G}{\delta \psi^{*}}&= \Big(\sum_{m=0}^{l} (-1)^{m} \partial_{x}^{m} \frac{\partial}{\partial(\partial_{x}^{m}\psi^{*})}\Big)   \Big[\frac{|\psi|^{2}}{2}\Big(\sum_{n=0}^{l} {l\choose n} (\partial_{x}^{l-n}\psi)  (\partial_{x}^{n} \psi^{*}) \Big) \Big]\\
&=\Big(\frac{\partial}{\partial \psi^{*}}+\sum_{m=1}^{l} (-1)^{m} \partial_{x}^{m} \frac{\partial}{\partial(\partial_{x}^{m}\psi^{*})}\Big) \Big[\frac{|\psi|^{2}}{2}\Big(\sum_{n=0}^{l} {l\choose n} (\partial_{x}^{l-n}\psi)  (\partial_{x}^{n} \psi^{*}) \Big) \Big]\\
&=\frac{\partial}{\partial \psi^{*}}\Big[\frac{|\psi|^{2}}{2}\Big(\sum_{n=0}^{l} {l\choose n} (\partial_{x}^{l-n}\psi)  (\partial_{x}^{n} \psi^{*}) \Big) \Big]\\
&+\sum_{m=1}^{l} (-1)^{m} \partial_{x}^{m} \frac{\partial}{\partial(\partial_{x}^{m}\psi^{*})} \Big[\frac{|\psi|^{2}}{2}\Big(\sum_{n=0}^{l} {l\choose n} (\partial_{x}^{l-n}\psi)  (\partial_{x}^{n} \psi^{*}) \Big) \Big]\\
&=\frac{\psi}{2}\partial_{x}^{l}(|\psi|^{2}) + \frac{1}{2} \Big[\sum_{m=0}^{l} (-1)^{m}{l\choose m} \partial_{x}^{m} \Big(|\psi|^{2}(\partial_{x}^{l-m}\psi) \Big) \Big],
\end{split}
\end{equation*}
\end{widetext}

which gives

\begin{widetext}
\begin{equation}
\frac{\delta G}{\delta \psi^{*}}=\frac{\psi}{2}\partial_{x}^{l}(|\psi|^{2}) + \frac{1}{2}\Big[\sum_{m=0}^{l} (-1)^{m}{l\choose m} \sum_{p=0}^{m}\Big({m\choose p} (\partial_{x}^{p}|\psi|^{2}) (\partial_{x}^{l-p} \psi )\Big) \Big].
\label{eq:1}
\end{equation}
\end{widetext}

Let us fix $p$ and considering the sum over $m$, look at the the second term on the right hand side in square brackets of the above Eq.(\ref{eq:1}), which becomes $\sum_{m=p}^{l} (-1)^{m}{l\choose m}{m\choose p} \partial_{x}^{p}|\psi|^{2} \partial_{x}^{l-p}\psi $. Notice the lower sum is from $p$ and not zero. This is necessary, because to have a term with $p$-th derivative, the minimum necessary value for the upper limit $'m'$ of the inner sum in the above equation is $p$. Also note that $p\leq m$. Let us introduce $q=m-p$. So we have,

\begin{widetext}
\begin{equation*}
\begin{split}
\sum_{m=p}^{l} (-1)^{m}{l\choose m}{m\choose p} (\partial_{x}^{p}|\psi|^{2}) (\partial_{x}^{l-p}\psi)&=\sum_{q=0}^{l-p} (-1)^{q+p}{l\choose q+p}{q+p\choose p} (\partial_{x}^{p}|\psi|^{2}) (\partial_{x}^{l-p}\psi)\\
&= \sum_{q=0}^{l-p} (-1)^{q+p}{l\choose p}{l-p\choose q} (\partial_{x}^{p}|\psi|^{2}) (\partial_{x}^{l-p}\psi)\\
&=(\partial_{x}^{p}|\psi|^{2}) (\partial_{x}^{l-p}\psi) (-1)^{p}{l\choose p}\sum_{q=0}^{l-p} (-1)^{q}{l-p\choose q},
\end{split}
\end{equation*}
\end{widetext}

where we have used the property ${l\choose q+p}{q+p\choose p}={l\choose p}{l-p\choose q}$. It is easy to see that $\sum_{q=0}^{l-p} (-1)^{q}{l-p\choose q}$ is nothing but the expansion of $(1-1)^{l-p}$ , which is zero, except for the case when $p=l$, for which, from Eq.(\ref{eq:1}) we can see that, $m=l$. So, we can conclude that the term $\frac{1}{2}\Big[\sum_{m=0}^{l} (-1)^{m}{l\choose m} \sum_{p=0}^{m}\Big({m\choose p} (\partial_{x}^{p}|\psi|^{2}) (\partial_{x}^{l-p} \psi )\Big) \Big]$ on the right hand side of Eq.(\ref{eq:1}) is zero, except when $p=m=l$. So, we need only consider the case where $p=m=l$ in which case, Eq.(\ref{eq:1}) assumes the following form

\begin{widetext}
\begin{equation}
\begin{split}
\frac{\delta G}{\delta \psi^{*}}&=\frac{\psi}{2}\partial_{x}^{l}(|\psi|^{2}) + \frac{1}{2}\Big[(-1)^{l}\Big(\psi (\partial_{x}^{l}|\psi|^{2})\Big) \Big]\\
&=\Big(\frac{\psi}{2}\partial_{x}^{l}|\psi|^{2}\Big) + (-1)^{l}\Big(\frac{\psi}{2}\partial_{x}^{l}|\psi|^{2}\Big).
\end{split}
\label{eq:2}
\end{equation}
\end{widetext}

From this it is clear that $\frac{\delta}{\delta \psi^{*}}(\frac{|\psi|^{2}}{2} \partial _{x}^{l}|\psi|^{2})=\psi\partial_{x}^{l}(|\psi|^{2})$ only when $l$ is even and $\frac{\delta}{\delta \psi^{*}}(\frac{|\psi|^{2}}{2} \partial _{x}^{l}|\psi|^{2})=0$ if $l$ is odd. Since the microscopic expansion for the s-wave scattering invloves only even powers of $k$\cite{nonlocal_pethick, nonlocal_wang} and hence only even order derivatives of $|\psi|^{2}$, the energy functional obtained by us can be used to calculate the energy of the system where we want to consider corrections to the local GP equation.

\par
This result is very important in 
the sense that the energy functional is very straightforward to evaluate, simply by writing the correction terms to the local GP equation and multiplying them by $\psi^{*}/2$ as long as s-wave scattering prevails. Now that we have the energy functional for corrections to the local GP equation, we can use it to study the effects of the corrections on the BEC.

\section{Discussions}
In the present paper, we have considered the standard form of corrections to the local GP equation arising from an effective range expansion due to non-local s-wave scattering. This effective range expansion gives corrections to the local GP equation. The order of expansion at which such a modified GP equation is truncated depends on BEC parameters. These parameters may include the effective range of the inter-boson interaction potential, the average inter-particle separation, etc. These corrections should be taken into account as we approach the diluteness limit as shown by Fu \textit{et al.} These correction terms typically are associated with the length scales in a BEC. The kinetic term in the local GP equation is represented by a Laplacian. Since the first correction term to this local GP equation is also a Laplacian, it is the term which is popularly considered in literature which studies BEC beyond the Fermi pseudopotential. In other words, the correction term probes a length scale similar to the kinetic term apart from the coefficient. To probe smaller length scales, one needs to add higher order corrections to the GP equation. 

\par
In this paper, we have discussed corrections to the local GP energy functional arising due to corrections to the local GP equation. We have given a rigorous combinatorial proof of the same for taking into consideration non-local corrections up to any order in derivatives of $|\psi|^{2}$. The non-triviality of this correspondence is represented by the fact that such a correspondence is absent for correction terms with odd order derivatives of $|\psi|^{2}$. Since the correction terms, as shown by Collin \textit{et al.}, contain only even order derivatives and not odd order ones, our result would help write energy functional for corrections up to any order. In this sense, such a correspondence is special to symmetric s-wave scattering in BECs. This correspondence can be used to study the energetics of corrections to the GP equation, beyond the Fermi pseudopotential.

\begin{acknowledgments}
I would like to thank my thesis supervisor Dr. Arijit Bhattacharyay for the discussions and help provided to give structure to this work. I would also like to acknowledge the support provided by the Council of Scientific and Industrial Research(CSIR), India.
\end{acknowledgments}

\bibliography{references}

\def\germ{\frak} \def\scr{\cal} \ifx\documentclass\undefinedcs
  \def\bf{\fam\bffam\tenbf}\def\rm{\fam0\tenrm}\fi 
  \def\defaultdefine#1#2{\expandafter\ifx\csname#1\endcsname\relax
  \expandafter\def\csname#1\endcsname{#2}\fi} \defaultdefine{Bbb}{\bf}
  \defaultdefine{frak}{\bf} \defaultdefine{=}{\B} 
  \defaultdefine{mathfrak}{\frak} \defaultdefine{mathbb}{\bf}
  \defaultdefine{mathcal}{\cal}
  \defaultdefine{beth}{BETH}\defaultdefine{cal}{\bf} \def\bbfI{{\Bbb I}}
  \def\mbox{\hbox} \def\text{\hbox} \def\om{\omega} \def\Cal#1{{\bf #1}}
  \def\pcf{pcf} \defaultdefine{cf}{cf} \defaultdefine{reals}{{\Bbb R}}
  \defaultdefine{real}{{\Bbb R}} \def\restriction{{|}} \def\club{CLUB}
  \def\w{\omega} \def\exist{\exists} \def\se{{\germ se}} \def\bb{{\bf b}}
  \def\equivalence{\equiv} \let\lt< \let\gt>
\begin{thebibliography}{10}%
\makeatletter
\providecommand \@ifxundefined [1]{%
 \ifx #1\undefined \expandafter \@firstoftwo
 \else \expandafter \@secondoftwo
\fi
}%
\providecommand \@ifnum [1]{%
 \ifnum #1\expandafter \@firstoftwo
 \else \expandafter \@secondoftwo
\fi
}%
\providecommand \enquote [1]{``#1''}%
\providecommand \bibnamefont  [1]{#1}%
\providecommand \bibfnamefont [1]{#1}%
\providecommand \citenamefont [1]{#1}%
\providecommand\href[0]{\@sanitize\@href}%
\providecommand\@href[1]{\endgroup\@@startlink{#1}\endgroup\@@href}%
\providecommand\@@href[1]{#1\@@endlink}%
\providecommand \@sanitize [0]{\begingroup\catcode`\&12\catcode`\#12\relax}%
\@ifxundefined \pdfoutput {\@firstoftwo}{%
 \@ifnum{\z@=\pdfoutput}{\@firstoftwo}{\@secondoftwo}%
}{%
 \providecommand\@@startlink[1]{\leavevmode\special{html:<a href="#1">}}%
 \providecommand\@@endlink[0]{\special{html:</a>}}%
}{%
 \providecommand\@@startlink[1]{%
  \leavevmode
  \pdfstartlink
   attr{/Border[0 0 1 ]/H/I/C[0 1 1]}%
   user{/Subtype/Link/A<</Type/Action/S/URI/URI(#1)>>}%
  \relax
 }%
 \providecommand\@@endlink[0]{\pdfendlink}%
}%
\providecommand \url  [0]{\begingroup\@sanitize \@url }%
\providecommand \@url [1]{\endgroup\@href {#1}{\urlprefix}}%
\providecommand \urlprefix [0]{URL }%
\providecommand \Eprint[0]{\href }%
\@ifxundefined \urlstyle {%
  \providecommand \doi [1]{doi:\discretionary{}{}{}#1}%
}{%
  \providecommand \doi [0]{doi:\discretionary{}{}{}\begingroup
  \urlstyle{rm}\Url }%
}%
\providecommand \doibase [0]{http://dx.doi.org/}%
\providecommand \Doi[1]{\href{\doibase#1}}%
\providecommand \bibAnnote [3]{%
  \BibitemShut{#1}%
  \begin{quotation}\noindent
    \textsc{Key:}\ #2\\\textsc{Annotation:}\ #3%
  \end{quotation}%
}%
\providecommand \bibAnnoteFile [2]{%
  \IfFileExists{#2}{\bibAnnote {#1} {#2} {\input{#2}}}{}%
}%
\providecommand \typeout [0]{\immediate \write \m@ne }%
\providecommand \selectlanguage [0]{\@gobble}%
\providecommand \bibinfo [0]{\@secondoftwo}%
\providecommand \bibfield [0]{\@secondoftwo}%
\providecommand \translation [1]{[#1]}%
\providecommand \BibitemOpen[0]{}%
\providecommand \bibitemStop [0]{}%
\providecommand \bibitemNoStop [0]{.\EOS\space}%
\providecommand \EOS [0]{\spacefactor3000\relax}%
\providecommand \BibitemShut [1]{\csname bibitem#1\endcsname}%
\bibitem{atomicbec_ketterle}%
  \BibitemOpen
  \bibfield{author}{%
  \bibinfo {author} {\bibfnamefont{W.~K.}\ \bibnamefont{et~al.}},\ }%
  \bibfield{journal}{%
  \bibinfo {journal} {Phys. Rev. Lett.}\ }%
  \textbf{\bibinfo {volume} {75, 3969}} (\bibinfo {year} {1995})%
  \bibAnnoteFile{NoStop}{atomicbec_ketterle}%
\bibitem{pethick}%
  \BibitemOpen
  \bibfield{author}{%
  \bibinfo {author} {\bibfnamefont{C.}~\bibnamefont{Pethick}}\ and\ \bibinfo
  {author} {\bibfnamefont{H.}~\bibnamefont{Smith}},\ }%
  \emph{\bibinfo {title} {Bose-Einstein Condensation in Dilute Gases}}\
  (\bibinfo {publisher} {Cambridge University Press},\ \bibinfo {year} {2001})%
  \bibAnnoteFile{NoStop}{pethick}%
\bibitem{gp_gross}%
  \BibitemOpen
  \bibfield{author}{%
  \bibinfo {author} {\bibfnamefont{E.~P.}\ \bibnamefont{Gross}},\ }%
  \bibfield{journal}{%
  \bibinfo {journal} {Annals of Physics}\ }%
  \textbf{\bibinfo {volume} {9}},\ \bibinfo {pages} {292} (\bibinfo {year}
  {1960})%
  \bibAnnoteFile{NoStop}{gp_gross}%
\bibitem{becreview_leggett}%
  \BibitemOpen
  \bibfield{author}{%
  \bibinfo {author} {\bibfnamefont{A.~J.}\ \bibnamefont{Leggett}},\ }%
  \bibfield{journal}{%
  \bibinfo {journal} {Rev. Mod. Phys.}\ }%
  \textbf{\bibinfo {volume} {73}},\ \bibinfo {pages} {307} (\bibinfo {month}
  {Apr}\ \bibinfo {year} {2001})%
  \bibAnnoteFile{NoStop}{becreview_leggett}%
\bibitem{excitations_mewes}%
  \BibitemOpen
  \bibfield{author}{%
  \bibinfo {author} {\bibfnamefont{M.-O.}\ \bibnamefont{Mewes}}, \bibinfo
  {author} {\bibfnamefont{M.}~\bibnamefont{Andrews}}, \bibinfo {author}
  {\bibfnamefont{N.}~\bibnamefont{Van~Druten}}, \bibinfo {author}
  {\bibfnamefont{D.}~\bibnamefont{Kurn}}, \bibinfo {author}
  {\bibfnamefont{D.}~\bibnamefont{Durfee}}, \bibinfo {author}
  {\bibfnamefont{C.}~\bibnamefont{Townsend}},\ and\ \bibinfo {author}
  {\bibfnamefont{W.}~\bibnamefont{Ketterle}},\ }%
  \bibfield{journal}{%
  \bibinfo {journal} {Phys. Rev. Lett.}\ }%
  \textbf{\bibinfo {volume} {77}},\ \bibinfo {pages} {988} (\bibinfo {year}
  {1996})%
  \bibAnnoteFile{NoStop}{excitations_mewes}%
\bibitem{excitations_stringari}%
  \BibitemOpen
  \bibfield{author}{%
  \bibinfo {author} {\bibfnamefont{S.}~\bibnamefont{Stringari}},\ }%
  \bibfield{journal}{%
  \bibinfo {journal} {Phys. Rev. Lett.}\ }%
  \textbf{\bibinfo {volume} {77}},\ \bibinfo {pages} {2360} (\bibinfo {year}
  {1996})%
  \bibAnnoteFile{NoStop}{excitations_stringari}%
\bibitem{solitons_jackson}%
  \BibitemOpen
  \bibfield{author}{%
  \bibinfo {author} {\bibfnamefont{A.}~\bibnamefont{Jackson}}, \bibinfo
  {author} {\bibfnamefont{G.}~\bibnamefont{Kavoulakis}},\ and\ \bibinfo
  {author} {\bibfnamefont{C.~J.}\ \bibnamefont{Pethick}},\ }%
  \bibfield{journal}{%
  \bibinfo {journal} {Phys. Rev. A}\ }%
  \textbf{\bibinfo {volume} {58}},\ \bibinfo {pages} {2417} (\bibinfo {year}
  {1998})%
  \bibAnnoteFile{NoStop}{solitons_jackson}%
\bibitem{solitons_burger}%
  \BibitemOpen
  \bibfield{author}{%
  \bibinfo {author} {\bibfnamefont{S.}~\bibnamefont{Burger}}, \bibinfo {author}
  {\bibfnamefont{K.}~\bibnamefont{Bongs}}, \bibinfo {author}
  {\bibfnamefont{S.}~\bibnamefont{Dettmer}}, \bibinfo {author}
  {\bibfnamefont{W.}~\bibnamefont{Ertmer}}, \bibinfo {author}
  {\bibfnamefont{K.}~\bibnamefont{Sengstock}}, \bibinfo {author}
  {\bibfnamefont{A.}~\bibnamefont{Sanpera}}, \bibinfo {author}
  {\bibfnamefont{G.}~\bibnamefont{Shlyapnikov}},\ and\ \bibinfo {author}
  {\bibfnamefont{M.}~\bibnamefont{Lewenstein}},\ }%
  \bibfield{journal}{%
  \bibinfo {journal} {Phys. Rev. Lett.}\ }%
  \textbf{\bibinfo {volume} {83}},\ \bibinfo {pages} {5198} (\bibinfo {year}
  {1999})%
  \bibAnnoteFile{NoStop}{solitons_burger}%
\bibitem{solitons_abdullaev}%
  \BibitemOpen
  \bibfield{author}{%
  \bibinfo {author} {\bibfnamefont{F.~K.}\ \bibnamefont{Abdullaev}}\ and\
  \bibinfo {author} {\bibfnamefont{R.}~\bibnamefont{Galimzyanov}},\ }%
  \bibfield{journal}{%
  \bibinfo {journal} {Journal of Physics B: Atomic, Molecular and Optical
  Physics}\ }%
  \textbf{\bibinfo {volume} {36}},\ \bibinfo {pages} {1099} (\bibinfo {year}
  {2003})%
  \bibAnnoteFile{NoStop}{solitons_abdullaev}%
\bibitem{surface_khawaja}%
  \BibitemOpen
  \bibfield{author}{%
  \bibinfo {author} {\bibfnamefont{U.}~\bibnamefont{Al~Khawaja}}, \bibinfo
  {author} {\bibfnamefont{C.}~\bibnamefont{Pethick}},\ and\ \bibinfo {author}
  {\bibfnamefont{H.}~\bibnamefont{Smith}},\ }%
  \bibfield{journal}{%
  \bibinfo {journal} {Phys. Rev. A}\ }%
  \textbf{\bibinfo {volume} {60}},\ \bibinfo {pages} {1507} (\bibinfo {year}
  {1999})%
  \bibAnnoteFile{NoStop}{surface_khawaja}%
\bibitem{surface_onofrio}%
  \BibitemOpen
  \bibfield{author}{%
  \bibinfo {author} {\bibfnamefont{R.}~\bibnamefont{Onofrio}}, \bibinfo
  {author} {\bibfnamefont{D.}~\bibnamefont{Durfee}}, \bibinfo {author}
  {\bibfnamefont{C.}~\bibnamefont{Raman}}, \bibinfo {author}
  {\bibfnamefont{M.}~\bibnamefont{K{\"o}hl}}, \bibinfo {author}
  {\bibfnamefont{C.}~\bibnamefont{Kuklewicz}},\ and\ \bibinfo {author}
  {\bibfnamefont{W.}~\bibnamefont{Ketterle}},\ }%
  \bibfield{journal}{%
  \bibinfo {journal} {Phys. Rev. Lett.}\ }%
  \textbf{\bibinfo {volume} {84}},\ \bibinfo {pages} {810} (\bibinfo {year}
  {2000})%
  \bibAnnoteFile{NoStop}{surface_onofrio}%
\bibitem{vortex_pitaevskii}%
  \BibitemOpen
  \bibfield{author}{%
  \bibinfo {author} {\bibfnamefont{L.}~\bibnamefont{Pitaevskii}},\ }%
  \bibfield{journal}{%
  \bibinfo {journal} {Sov. Phys. JETP}\ }%
  \textbf{\bibinfo {volume} {13}},\ \bibinfo {pages} {451} (\bibinfo {year}
  {1961})%
  \bibAnnoteFile{NoStop}{vortex_pitaevskii}%
\bibitem{vortices_fetter}%
  \BibitemOpen
  \bibfield{author}{%
  \bibinfo {author} {\bibfnamefont{A.~L.}\ \bibnamefont{Fetter}}\ and\ \bibinfo
  {author} {\bibfnamefont{A.~A.}\ \bibnamefont{Svidzinsky}},\ }%
  \bibfield{journal}{%
  \bibinfo {journal} {Journal of Physics: Condensed Matter}\ }%
  \textbf{\bibinfo {volume} {13}},\ \bibinfo {pages} {R135} (\bibinfo {year}
  {2001})%
  \bibAnnoteFile{NoStop}{vortices_fetter}%
\bibitem{vortices_lundh}%
  \BibitemOpen
  \bibfield{author}{%
  \bibinfo {author} {\bibfnamefont{E.}~\bibnamefont{Lundh}}, \bibinfo {author}
  {\bibfnamefont{C.~J.}\ \bibnamefont{Pethick}},\ and\ \bibinfo {author}
  {\bibfnamefont{H.}~\bibnamefont{Smith}},\ }%
  \bibfield{journal}{%
  \bibinfo {journal} {Phys. Rev. A}\ }%
  \textbf{\bibinfo {volume} {58}},\ \bibinfo {pages} {4816} (\bibinfo {year}
  {1998})%
  \bibAnnoteFile{NoStop}{vortices_lundh}%
\bibitem{ps}%
  \BibitemOpen
  \bibfield{author}{%
  \bibinfo {author} {\bibfnamefont{L.}~\bibnamefont{Pitaevskii}}\ and\ \bibinfo
  {author} {\bibfnamefont{S.}~\bibnamefont{Stringari}},\ }%
  \emph{\bibinfo {title} {Bose-Einstein Condensation}}\ (\bibinfo {publisher}
  {Oxford Science Publications},\ \bibinfo {year} {2003})%
  \bibAnnoteFile{NoStop}{ps}%
\bibitem{nonlocal_wang}%
  \BibitemOpen
  \bibfield{author}{%
  \bibinfo {author} {\bibfnamefont{H.}~\bibnamefont{Fu}}, \bibinfo {author}
  {\bibfnamefont{Y.}~\bibnamefont{Wang}},\ and\ \bibinfo {author}
  {\bibfnamefont{B.}~\bibnamefont{Gao}},\ }%
  \bibfield{journal}{%
  \bibinfo {journal} {Phys. Rev. A}\ }%
  \textbf{\bibinfo {volume} {67}},\ \bibinfo {pages} {053612} (\bibinfo {year}
  {2003})%
  \bibAnnoteFile{NoStop}{nonlocal_wang}%
\bibitem{nonlocal_pethick}%
  \BibitemOpen
  \bibfield{author}{%
  \bibinfo {author} {\bibfnamefont{A.}~\bibnamefont{Collin}}, \bibinfo {author}
  {\bibfnamefont{P.}~\bibnamefont{Massignan}},\ and\ \bibinfo {author}
  {\bibfnamefont{C.}~\bibnamefont{Pethick}},\ }%
  \bibfield{journal}{%
  \bibinfo {journal} {Physical Review A}\ }%
  \textbf{\bibinfo {volume} {75}},\ \bibinfo {pages} {013615} (\bibinfo {year}
  {2007})%
  \bibAnnoteFile{NoStop}{nonlocal_pethick}%
\bibitem{nonlocal_ketterle}%
  \BibitemOpen
  \bibfield{author}{%
  \bibinfo {author} {\bibfnamefont{H.}~\bibnamefont{Veksler}}, \bibinfo
  {author} {\bibfnamefont{S.}~\bibnamefont{Fishman}},\ and\ \bibinfo {author}
  {\bibfnamefont{W.}~\bibnamefont{Ketterle}},\ }%
  \bibfield{journal}{%
  \bibinfo {journal} {Physical Review A}\ }%
  \textbf{\bibinfo {volume} {90}},\ \bibinfo {pages} {023620} (\bibinfo {year}
  {2014})%
  \bibAnnoteFile{NoStop}{nonlocal_ketterle}%
\end{thebibliography}%
\end{document}